\documentclass[12pt]{article}
\usepackage{amsmath,amssymb,amsopn,amsthm,times,cite}

\title{The uses of Connes and\\ Kreimer's algebraic formulation of\\
renormalization theory}

\author{
H\'ector Figueroa* and
Jos\'e M. Gracia-Bond\'{\i}a\dag,\ddag
\\[1pc]
*\,Department of Mathematics,\\
\dag\,Department of Physics,\\
Universidad de Costa Rica, 2060 San Pedro, Costa Rica\\
and\\
\ddag\,Department of Theoretical Physics, Universidad de Zaragoza,\\
50009 Zaragoza, Spain}

\scrollmode

\advance\topmargin by -\headheight
\advance\topmargin by -\headsep
\evensidemargin=\oddsidemargin
\theoremstyle{plain}
\newtheorem{thm}{Theorem}           
\newtheorem{prop}[thm]{Proposition} 

\theoremstyle{definition}
\newtheorem{defn}{Definition}       

\newcommand{\Cc}{\mathcal{C}}      
\newcommand{\D}{\mathcal{D}}       
\newcommand{\Dl}{\Delta}           
\newcommand{\dl}{\delta}           
\newcommand{\eps}{\varepsilon}     
\newcommand{\F}{\mathcal{F}}       
\newcommand{\FF}{\mathbb{F}}       
\newcommand{\Ga}{\Gamma}           
\newcommand{\ga}{\gamma}           
\renewcommand{\H}{\mathcal{H}}     
\DeclareMathOperator{\Hom}{Hom}    
\newcommand{\id}{\mathrm{id}}      
\renewcommand{\L}{\mathcal{L}}     
\newcommand{\la}{\lambda}          
\newcommand{\longto}{\mathop{\longrightarrow}\limits} 
\newcommand{\Om}{\Omega}           
\newcommand{\ox}{\otimes}          
\newcommand{\oxyox}{\otimes\cdots\otimes} 
\newcommand{\row}[3]{{#1}_{#2},\dots,{#1}_{#3}} 
\newcommand{\sepword}[1]{\qquad\hbox{#1}\quad} 
\newcommand{\set}[1]{\{\,#1\,\}}   
\newcommand{\sg}{\sigma}           
\newcommand{\Th}{\Theta}           
\newcommand{\tsum}{\mathop{\textstyle\sum}\nolimits} 
\newcommand{\V}{\mathcal{V}}       
\newcommand{\7}{\dagger}           
\renewcommand{\:}{\colon}          
\def\<#1,#2>{\langle#1,#2\rangle}  

\makeatletter
\def\section{\@startsection{section}{1}{\z@}{-3.5ex plus -1ex minus
      -.2ex}{2.3ex plus .2ex}{\large\bf}}
\def\subsection{\@startsection{subsection}{2}{\z@}{-3.25ex plus -1ex
      minus -.2ex}{1.5ex plus .2ex}{\normalsize\bf}}
\makeatother

\newcommand{\hideqed}{\renewcommand{\qed}{}} 

\begin{document}

\maketitle

\begin{abstract}
We show how, modulo the distinction between the antipode and the
``twisted'' or ``renormalized'' antipode, Connes and Kreimer's
algebraic paradigm trivializes the proofs of equivalence of the
(corrected) Dyson--Salam, Bogoliubov--Parasiuk--Hepp and Zimmermann
procedures for renormalizing Feynman amplitudes. We discuss the
outlook for a parallel simplification of computations in quantum field
theory, stemming from the same algebraic approach.
\end{abstract}

\medskip



\section{Introduction}

The present authors have dealt in a Hopf algebraic context with the relation
between the Dyson--Salam, Bogoliubov--Parasiuk and Zimmermann renormalization
schemes in Ref.~1. This we did using the algebra of rooted trees
$H_R$ (Ref.~2) as a proxy for the complexities of the
combinatorics of Feynman graphs in renormalization.

The point of Ref.~1 was that the differences of the diverse
schemes could largely be tracked down to avatars of the convolution operation
in spaces of homomorphisms of Hopf algebra. This was illustrated by the
apparently simple-minded, but tremendously effective, computation of antipode
images in $H_R$ by means of the convolution geometric series.

In this paper we return on the subject, now in terms of Hopf algebras of the
Feynman graphs themselves.

It must be acknowledged that the deeply conceptual approach by Connes
and Kreimer till now has failed to impress many physicists who
practise the renormalization theory. For that approach to become
mainstream, it should be shown to simplify both \textit{proofs} and
\textit{calculations} in perturbative renormalization theory.

The proof given by Zimmermann${}^{3}$ of the equivalence of
his forest formula and Bogo\-liubov and Parasiuk's scheme has a reputation of
difficulty. The Dyson--Salam scheme has been thoroughly analyzed and
purged of difficulties in Ref.~4. But the proof of equivalence
between the forest and the (corrected) Dyson--Salam formulae given
in Ref.~4 is anything but simple.

In contrast, here we show that the equivalence between all the
aforementioned schemes boils down to downright basic facts in Hopf
algebra theory. So basic in fact, that we can only conclude that the
combinatorics of perturbative renormalization finds its definitive
expression in Hopf algebraic terms. Not only the equivalence proofs
result from uniqueness of the Hopf antipode, but we are able to show
that the Dyson--Salam scheme corresponds \textit{identically} to the
convolution geometric series. All our arguments are elementary and
short, much more so, by the way, than in Ref.~1: in the
previous paper, we did not quite see the forest for the trees.

At the end of the paper we briefly review the perspectives for the
Connes--Kreimer method to simplify other theorems and
\textit{computations} in renormalization, by means of reduction to the
case of elements that are primitive (i.e., without subdivergences).
Here the situation is more mixed; still, the hope remains that
sizeable simplifications can be gleaned from the algebraic approach.

\section{Bialgebras of graphs}

The basics of graded bialgebra theory are recalled in the Appendix. {}From now
on, we assume the reader is familiar with them.

Bialgebras of Feynman graphs, encoding the combinatorics of renormalization,
were introduced by Connes and Kreimer in Ref.~5. The precise
definition we use in this paper was first given in Ref.~6. To fix
ideas, and whenever an example is given, we think of the (massless)
$\varphi^4_4$ scalar model. Nevertheless, the constructions hold in any given
quantum field theory, such as the $\varphi^3_6$ model considered in
Ref.~5.

We recall that a \textit{graph} or diagram $\Ga$ of the theory is specified by
a set $\V(\Ga)$ of \textit{vertices} and a set $\L(\Ga)$ of \textit{lines}
(propagators) among them; \textit{external} lines are attached to only one
vertex each, \textit{internal} lines to two. Diagrams with no external lines
will not be taken into account ---and in $\varphi^4_4$ theory only graphs with
an even number of external lines are to be found. Also tadpole diagrams, in
which a line connects a vertex to itself, are excluded in this paper.

Given a graph $\Ga$, a \textit{subdiagram} $\ga$ of $\Ga$ is specified by a
subset of at least two elements of $\V(\Ga)$ and a subset of the lines that
join these vertices in $\Ga$. By exception, the empty subset $\emptyset$ will
be admitted as a subdiagram of $\Ga$. As well as $\Ga$ itself. Clearly, the
external lines for a subdiagram $\ga$ include not only a subset of original
incident lines, but some internal lines of $\Ga$ not included in $\ga$. The
connected pieces of~$\Ga$ are the maximal connected subdiagrams. A diagram is
\textit{proper} (or 1PI) when the number of its connected pieces would not
increase on the removal of a single internal line; otherwise it is called
\textit{improper}. An improper graph is the union of proper components plus
subdiagrams containing a single line.

A \textit{subgraph} of a proper graph is a subdiagram that contains all the
elements of $\L(\Ga)$ joining its vertices in the whole graph; as such, it is
determined solely by the vertices. When a subdiagram contains several
connected pieces, each one of them being a subgraph, we still call it a
subgraph. A subgraph of an improper graph $\Ga$, distinct from $\Ga$ itself,
is a proper subdiagram each of whose components is a subgraph with respect to
the proper components of~$\Ga$.

We write $\ga \subseteq \Ga$ if and only if $\ga$ is a subgraph of~$\Ga$ as
defined (not just a subdiagram): this is the really important concept for us.
For renormalization in configuration space,${}^{\7}$ it is more
convenient to deal with subgraphs than with more general subdiagrams.
Zimmermann showed long ago that only subtractions corresponding to subgraphs
need be used,${}^{8}$ and this dispenses us from dealing with
subdiagrams that are not subgraphs.

Two subgraphs $\ga_1,\ga_2$ of $\Ga$ are said to be
\textit{nonoverlapping} when $\ga_1 \cap \ga_2 = \emptyset$ or
$\ga_1 \subseteq \ga_2$ or $\ga_2 \subseteq \ga_1$; otherwise they are
overlapping. Given $\ga \subseteq \Ga$, the quotient graph or cograph
$\Ga/\ga$ (reduced graph in Zimmermann's parlance) is defined by shrinking
$\ga$ in $\Ga$ to a point, that is to say, $\ga$ (bereft of its external
lines) is considered as a vertex of $\Ga$, and all the lines in $\Ga$ not
belonging to $\ga$ belong to $\Ga/\ga$. This is modified in the obvious way
when $\ga$ represents a propagator correction. The graphs $\Ga$ and
$\Ga/\ga$ have the same external structure. A nonempty $\Ga/\ga$ will be
proper iff $\Ga$ is proper ---the situation considered
in Ref.~5.

Now, the bialgebra $\H$ is defined as the polynomial algebra generated by
the empty set $\emptyset$ and the connected Feynman graphs that are
(superficially) divergent and/or have (superficially) divergent subgraphs
(renormalization parts in Zimmermann's parlance), with set union as the
product operation (hence $\emptyset$ is the unit element $1\in\H$). The
counit is given by $\eps(\Ga) := 0$ on any generator, except
$\eps(\emptyset) = 1$.

The really telling operation is the coproduct $\Dl: \H \to \H \ox \H$;
as it is to be a homomorphism of the algebra structure, we need only
define it on connected diagrams. By definition, the coproduct of $\Ga$
is given by
\begin{equation}
\Dl \Ga\, := \Ga \ox 1 + 1 \ox \Ga
+ \sum_{\emptyset \varsubsetneq \ga \varsubsetneq \Ga} \ga \ox \Ga/\ga\, =
   \sum_{\emptyset \subseteq \ga \subseteq \Ga} \ga \ox \Ga/\ga.
\label{eq:coprod} 
\end{equation}
The sum is over all divergent, proper, not necessarily connected
subgraphs of $\Ga$, such that \textit{each piece} is divergent,
including (and then with the possible exception of, as $\Ga$ need not
be divergent nor proper) the empty set and $\Ga$ itself. We put $\Ga/\Ga=1$.
When appropriate, the sum runs also over different types of local counterterms
associated to $\ga$ (see Refs.~5,9);
this is not needed in our example model.

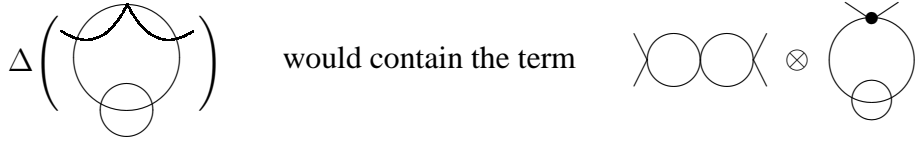
\begin{figure}[ht]
\begin{center}
\vspace{2pc}
\parbox{5pc}{
\begin{picture}(50,18)
\put(-20,5){$\Delta$}
\put(-10,5){$\Biggl($}
\put(25,10){\circle{40}}
\put(25,-10){\circle{20}}
\qbezier(25,30)(35,10)(50,20)
\qbezier(25,30)(15,10)(0,20)
\put(52,5){$\Biggr)$}
\end{picture}
}\qquad
would contain the term
\qquad
\parbox{7pc}{
\begin{picture}(80,10)
\put(0,5){\line(-1,2){5}}
\put(0,5){\line(-1,-2){5}}
\put(10,5){\circle{20}}
\put(30.2,5){\circle{20}}
\put(40.2,5){\line(1,2){5}}
\put(40.2,5){\line(1,-2){5}}
\put(52,2){$\otimes$}
\put(85,5){\circle{30}}
\put(85,-10){\circle{15}}
\put(85,20){\line(3,2){10}}
\put(85,20){\line(-3,2){10}}
\put(82,18){$\bullet$}
\end{picture}
}
\end{center}
\vspace{1pc}
\caption{Cograph which is a tadpole part}
\label{fg:tadpole-corr}
\vspace{1pc}
\end{figure}

It is natural to exclude the appearance of tadpole parts in $\Ga/\ga$, and
this we do hereafter. This tadpole-free condition was not used or remarked
in Ref.~6, nor in Ref.~10, which employs the same
definition. We show in Figure~\ref{fg:tadpole-corr} how this situation can
happen. The cograph corresponding to the ``bikini'' subgraph in the upper
part of the graph in Figure~\ref{fg:tadpole-corr} is a tadpole correction
and can be outlawed from the coproduct.

For the proof of the bialgebra properties of $\H$, we refer to Ref.~11; for
graphical examples of coproducts, see Ref.~6.

Actually $\H$ is a connected, graded bialgebra. Obvious grading operators
are available: if $\#(\Ga)$ denotes the number of vertices in $\Ga$ (i.e.,
the coupling order), then we define the degree of a generator (connected
element) $\Ga$ as $\nu(\Ga) := \#(\Ga) - 1$; the degree of a product is the
sum of the degrees of the factors. This grading is compatible with the
coproduct, and clearly scalars are the only degree~0 elements. Other
gradings are by the number $I(\Ga)$ of internal lines in $\Ga$ and by loop
number $\ell(\Ga) := I(\Ga) - \nu(\Ga)$. For the $\varphi^4_4$ model,
$\ell(\Ga) = \nu(\Ga) + 1$ for two-point graphs and $\ell(\Ga) = \nu(\Ga)$
for four-point graphs. A more relevant grading will emerge in the next
section.

Lurking in the background there is a character (i.e., multiplicative) map
$f$ (the ``Feynman rule'') of $\H$ into an algebra $V$ of Feynman
amplitudes: for instance, in dimensional regularization the character takes
values in a ring of Laurent series (with finite order poles) in the
regularization parameter. In physics, the Feynman rules are essentially
fixed by the interpretation of the theory, and thus one tends to identify
$\Ga$ with $f(\Ga)$.

\section{The importance of convolution}

Given a unital algebra $(A,m,u)$ and a counital coalgebra
$(C,\Dl,\eps)$ over $\FF$, the \textit{convolution} of two elements
$f,g$ of the vector space of $\FF$-linear maps $\Hom(C,A)$ is defined
as the map $f * g \in \Hom(C,A)$ given by the composition
$$
C \longto^\Dl C \ox C \longto^{f\ox g} A \ox A \longto^m A.
$$
In other words, $f * g = m (f \ox g) \Dl$. This product
turns $\Hom(C,A)$ into a unital algebra, where the unit is the map
$u \eps$, as is easily checked. In particular, linear endomorphisms of
a bialgebra can be convolved.

A bialgebra $H$ in which the identity map $\id_H$ is invertible under
convolution is called a \textit{Hopf algebra}, and its convolution
inverse $S$ is called the coinverse or \textit{antipode}; that is to
say, $\id_H * S = S * \id_H = u \eps$. The antipode is clearly unique.
It is known to be of order two for commutative bialgebras. Also, in a
commutative bialgebra $S$ is a homomorphism: $S(ab) = S(a)S(b)$.

In particular, if $\Dl(a) = \sum_j a'_j \ox  a''_j$, then
\begin{equation}
\eps(a) 1_H = u \eps(a) = m (S \ox \id) \Dl(a)
= \tsum_j S(a'_j) a''_j,
\label{eq:antipode}
\end{equation}
and likewise $\eps(a) 1_H = \sum_j a'_j S(a''_j)$. Since any left
inverse under convolution automatically equals any right inverse
provided both exist, any map $S$ satisfying \eqref{eq:antipode} is the
antipode.

The main outcome of equation~\eqref{eq:gr-coprod} in the Appendix for
connected graded bialgebras is that these \textit{are always Hopf}.
Indeed, as is done in Ref.~1, one can try to compute the antipode
$S\: H \to H$ by exploiting its very definition as the convolution
inverse of the identity in $H$, via a geometric series:
\begin{equation}
S := (\id)^{*-1} = (u \eps -(u \eps -\id))^{*-1}
      = u \eps + (u \eps -\id) + (u \eps -\id)^{*2}
      + \cdots
\label{eq:geom-series}
\end{equation}

\begin{prop}
Let $H$ be a connected, graded bialgebra, then the geometric series
expansion of $S(a)$ has at most $n+1$ terms when $a \in H^{(n)}$.
\end{prop}

\begin{proof}
If $a \in H^{(0)}$ the claim holds since $(u \eps - \id)1 = 0$.
Assume that the claim holds for the elements of $H^{(k)}$ when
$k \leq n-1$, and let $a \in H^{(n)}$; then by~\eqref{eq:gr-coprod}
\begin{align*}
(u \eps -\id)^{*(n+1)}(a)
&= (u \eps -\id) * (u \eps -\id)^{*n}(a)
\\
&= m [(u \eps -\id) \ox (u \eps -\id)^{*n}] \Dl(a)
\\
&= m [(u \eps -\id) \ox (u \eps -\id)^{*n}]
(a \ox 1 + 1 \ox a + \Dl'a).
\end{align*}
The first two terms vanish because $(u \eps - \id)1 = 0$. By the
induction hypothesis each of the summands of the third term are also
zero.
\end{proof}

As a corollary, connected graded bialgebras are always Hopf, with
antipode indeed given by the geometric series~\eqref{eq:geom-series}.
One of the advantages of this formulation is that we obtain fully
explicit formulae for $S$ from the coproduct.

\begin{prop}
If $a \in H^{(n)}$, $\Dl'(a) = \sum_{j_1} a'_{j_1} \ox a''_{j_1}$,
$\Dl'(a'_{j_1}) = \sum_{j_2} a'_{j_1j_2} \ox a''_{j_1j_2}$, and in
general $\Dl'(a'_{\row  j1k})
= \sum_{j_{k+1}} a'_{\row j1{k+1}} \ox a''_{\row j1{k+1}}$, then for
$1 \leq k\leq n -1$,
\begin{equation}
(u \eps -\id)^{*k+1}(a) = (-1)^{k+1} \sum_{\row j1k}
a'_{\row j1k} \, a''_{\row j1k} \cdots a''_{j_1 j_2} \, a''_{j_1}.
\label{eq:many-conv}
\end{equation}
\end{prop}

\begin{proof}
To abbreviate we use the notation $\sg := u\eps - \id$. Then
$\sg(a) = -a$ if $a \in H^{(n)}$ with $n \geq 1$, because then
$\eps(a) = 0$. Moreover,
\begin{align*}
\sg^{*2}(a) &= m (\sg \ox \sg) (a \ox 1 + 1 \ox a + \Dl'a)
\\
&= \sum_{j_1} \sg(a'_{j_1}) \, \sg(a''_{j_1})
   = \sum_{j_1} a'_{j_1} \, a''_{j_1},
\end{align*}
so the statement holds for $k = 1$. If the statement holds for~$\sg^{*k}$,
then
\begin{align*}
\sg^{*k+1}(a) &= m (\sg^{*k} \ox \sg) (a \ox 1 + 1 \ox a + \Dl'a)
\\
&= \sum_{j_1} \sg^{*k}(a'_{j_1}) \, \sg(a''_{j_1})
   = - \sum_{j_1} \sg^{*k}(a'_{j_1}) \, a''_{j_1}
\\
&=  (-1)^{k+1} \sum_{{\row j1k}}
a'_{\row  j1k} \, a''_{\row j1k} \cdots a''_{j_1 j_2} \, a''_{j_1},
\end{align*}
since, by the induction hypothesis,
$\sg^{*k}(a'_{j_1}) = (-1)^k \sum_{{\row j2k}}
a'_{j_1,\row j2k} \, a''_{j_1,\row j2k} \cdots a''_{j_1,j_2}$.
\end{proof}

If $\Dl'_i$ denotes the map $H^{\ox i} \to H^{\ox i+1}$ where $\Dl'$
is applied on the first tensor factor only, and $m_i$ the map
$H^{\ox(i+1)} \to H^{\ox i}$, where $m$ is applied on the first two
tensor factors only, then we can rewrite \eqref{eq:many-conv} as
\begin{equation}
(u \eps -\id)^{*k+1} = (-1)^{k+1} m\,m_2 \cdots
m_k \Dl'_k \cdots \Dl'_2 \Dl'.
\label{eq:many-conv-bis}
\end{equation}
By splitting the powers in the form $\sg^{*(k+1)} = \sg * \sg^k$, one
can obtain a twin formula of~\eqref{eq:many-conv}, on which the
coproduct is applied successively on the last tensor factor, instead
of on the first factor; that formula was given in Ref.~1.

\smallskip

While all the foregoing is happily elementary, the following important fact
must be registered: a new grading on $H$ has been obtained, defined simply
by declaring the degree of a generator $a$ as $k$ when $\sg^{*k}(a) \neq 0$
and $\sg^{*k+1}(a) = 0$. It is easily seen that this indeed defines a
grading $\dl$; the degree of a product is the sum of the degrees of the
factors.

For a connected diagram $\Ga \in \H$ this grading coincides with
the maximal length of a chain of subdivergences inside $\Ga$ (see
below), and in this context we call it \textit{depth}. We can
combine $\dl$ with the gradings $\#$ or $\ell$ to obtain a
bidegree: note that we have proved $\dl(\Ga) \leq\ \#(\Ga)$, for
any $\Ga$.

This grading by depth is the same $k$-primitivity grading already pondered in
the seminal paper${}^{12}$ and studied in the context
of~$H_R$ by Broadhurst and Kreimer.${}^{13}$ Here the concept is even
more pertinent, as the correlation between loop number and depth in field
theory is weaker than the correlation between the number of tree vertices and
$k$-primitivity in $H_R$: for instance, it is well known that in the
$\varphi^4_4$ model there are three 5-loop diagrams which are
(1-)primitive.${}^{14}$

We write $\H^{(k)}$ for the space of elements of primitivity degree $k$.

\smallskip

Anticipating the following discussion, note that there are other ways
to show that a connected graded bialgebra is a Hopf algebra. One can
take advantage of the equation $m (S\ox\id) \Dl(a) = 0$ whenever
$a \in H^{(n)}$ for $n \ge 1$, to introduce in the context the
Bogoliubov recursive formula:
\begin{equation}
S_B(a) := - a - \sum_{j} S_B(a'_j) a''_j,
\label{eq:Bogol-recur}
\end{equation}
if $\Dl'a = \sum_j a'_j \ox  a''_j$.

\begin{prop}
\label{pr:Bogol-antp}
If $H$ is a connected, graded bialgebra, then $S(a) = S_B(a)$.
\end{prop}

\begin{proof}
The statement holds, by a direct check, if $a \in H^{(1)}$. Assume
that $S(b) = S_B(b)$ whenever $b \in H^{(k)}$ with $k \leq n$, and let
$a \in H^{(n+1)}$. Then
\begin{align*}
S(a) &= \sg(a) + \sum_{i=1}^n \sg^{*i} * \sg(a)
    = -a + m \biggl( \sum_{i=1}^n \sg^{*i} \ox \sg \biggr) \Dl(a)
\\
&= -a + m \sum_{i=1}^n \sg^{*i} \ox \sg (a \ox 1 + 1 \ox a + \Dl'a)
\\
&= -a + \sum_j \sum_{i=1}^n \sg^{*i} (a'_j) \, \sg(a''_j)
  = -a - \sum_j \sum_{i=1}^n \sg^{*i} (a'_j) \, a''_j
\\
&=  -a - \sum_j S_B(a'_j) \,a''_j = S_B(a),
\end{align*}
where the penultimate equality uses the inductive hypothesis.
\end{proof}

Taking into account that we can also write
$S(a) = \sg(a) + \sum_{i=1}^n \sg * \sg^{*i}(a)$, it follows that the
twin formula
$$
S'_B(a) := - a - \sum_{j} a'_j S'_B(a''_j),
$$
also provides an expression for the antipode.

We record~\eqref{eq:Bogol-recur} in the language of the bialgebra $\H$
of graphs
\begin{equation}
S(\Ga) := - \Ga -
\sum_{\emptyset \varsubsetneq \ga \varsubsetneq \Ga} S(\ga) \, \Ga/\ga.
\label{eq:Bogol-recurgraph}
\end{equation}
For a primitive diagram, $S(\Ga) = -\Ga$.

\smallskip
Now it is time to reveal our strategy. Perhaps the main path-breaking
insight of Ref.~12 and subsequent papers by Kreimer and
coworkers is the introduction of the ``twisted antipode''. Let us  usher in
the other personages of this drama. There is a linear map $T: V \to V$,
which effects the subtraction of ultraviolet divergencies in each
renormalization scheme.

The twisted (or ``renormalized'') antipode $S_{T,f}$ is a map
$\H \to V$ defined by $S_{T,f}(\emptyset) = 1$; $S_{T,f} = T\circ
f\circ S$ for primitive diagrams, and then recursively:
$$
S_{T,f} \Ga = -[T\circ f]\Ga - T\biggl[
\sum_{\emptyset\varsubsetneq\ga\varsubsetneq\Ga} S_{T,f}(\ga) \, f(\Ga/\ga)
\biggr].
$$
In other words, $S_{T,f}$ is the map that produces the counterterms in
perturbative field theory. The Hopf algebra approach works most effectively
because in many cases $S_{T,f}$ is multiplicative; for that, it is not
necessary for $T$ to be an endomorphism of the algebra of amplitudes $V$, but
the following weaker condition${}^{5,15,16}$ is sufficient:
$$
T(hg) = T(T(h)g) + T(hT(g)) - T(h)T(g).
$$
This condition endows $\H$ with the structure of a Rota--Baxter algebra (see
Refs.~17,18); it is fulfilled in the BPHZ formalism and the dimensional
regularization scheme with minimal subtraction, for which the present paradigm
is most cleanly formulated.${}^{19}$

Finally, the renormalization map $R_{T,f}$ is given by
$$
R_{T,f} := S_{T,f} * f.
$$
In view of a previous remark, $R_{T,f}$ is also a homomorphism; compatibility
with the coproduct operation is given by its very definition as a convolution.

(In Epstein--Glaser renormalization, things are a bit more
complicated, since $S_{T,f}$ is not properly defined, and $R_{T,f}$
involves a map between two different spaces of Feynman amplitudes;
still, the homomorphism condition for $R_{T,f}$ can be enforced, and
$R_{T,f}$ is compatible with the Hopf algebra structure in a suitable
sense.)

In what follows, we shall assume that $S_{T,f}$ has been defined to be a
homomorphism, and we concentrate on the computation of $S$. According to
the dictum, Hopf algebras simplify combinatorics by reducing it to
algebra. Connes and Kreimer's algebraic approach to the renormalization
schemes separates neatly their combinatorics from the analytical
procedures and renders the first an essentially trivial application of Hopf
algebra.

\smallskip

Now, for the combinatorial aspect in renormalization theory, there are on
the market mainly the recursive formula by Bogoliubov, Zimmermann's forest
formula and the corrected Dyson--Salam formula. The last one is most
natural in the context of the primitivity grading. They just amount to
different ways to compute the antipode. It must be already evident that
the recursive formula by Bogoliubov corresponds to the definition of
$S_B$. Now, in order to prove the equivalence of two combinatorial
schemes, it is enough to prove that both yield the antipode, either
directly, as in the proof of Proposition~3, or by using the uniqueness of
the antipode. This we systematically proceed to do in the sequel. That
the coming proofs are all short, utterly simple, or decidedly trivial, is
our main point and asset.

It turns out, and this is perhaps the most illuminating result, that the
Dyson--Salam scheme corresponds identically (i.e., without need of
further cancellations) to the geometric series formula.

\section{Convolution and the Dyson--Salam formula}

The present framework applies to proper and improper graphs. For brevity, in
what follows we concentrate on proper graphs.

\begin{defn}
A \textit{chain} $\Cc$ of a proper, connected graph $\Ga$ is a
sequence $\emptyset\varsubsetneq\ga_1\varsubsetneq\ga_2
\varsubsetneq\cdots\varsubsetneq\ga_k \varsubsetneq\Ga$ of proper,
divergent, \textit{not necessarily connected} subgraphs of $\Ga$.  We
denote by $C(\Ga)$ the set of chains of $\Ga$.  The \textit{length} of
a chain $\Cc$ is the number $l(\Cc) = k + 1 =: |\Cc| + 1$, and we
write $\Om(\Cc) := \ga_1\,(\ga_2/\ga_1)\dots(\ga_{k-1}/\ga_k)\,(\Ga/\ga_k)$.

With this notation we can define the antipode as follows:
\begin{equation}
S_{DS}(\Ga) := \sum_{\Cc \in C(\Ga)} (-1)^{l(\Cc)} \Om(\Cc).
\label{eq:ds-antip}
\end{equation}
\end{defn}

This definition corresponds, on the one hand, to the correct version of
the Dyson--Salam formula for renormalization. On the other hand,
formula~\eqref{eq:ds-antip} is totally analogous to the explicit
expression for the antipode given by Schmitt for his incidence Hopf
algebras.${}^{20}$

\begin{prop}
\label{pr:DS-antp}
$S_{DS}$ so defined is an antipode for $\H$.
\end{prop}

\begin{proof}
We prove that $S_{DS}$ is an inverse, under convolution, of $\id$. By
definition,
\begin{align*}
S_{DS} * \id (\Ga)
&= \sum_{\ga \subseteq \Ga} S_{DS}(\ga) \, \Ga/\ga
= S_{DS}(\Ga) + \sum_{\emptyset \subseteq  \ga \varsubsetneq \Ga}
S_{DS}(\ga) \, \Ga/\ga
\\
&= S_{DS}(\Ga) + \sum_{\emptyset \subseteq  \ga \varsubsetneq \Ga}
\sum_{\D \in C(\ga)}  (-1)^{l(\D)} \Om(\D) \, \Ga/\ga.
\end{align*}

Now, if $\D \in C(\ga)$, say $\D = \{\row{\ga}1k\}$, then
$\Cc = \{\row{\ga}1k,\ga\} \in C(\Ga)$. Moreover,
\begin{equation}
\Om(\Cc) = \Om(\D) \, \Ga/\ga, \sepword{and}\quad
l(\Cc) = l(\D) + 1.
\label{eq:chain-length}
\end{equation}
On the other hand, given a chain $\Cc = \{\row{\ga}1n\} \in C(\Ga)$,
then $\D = \{\row {\ga}1{n-1}\} \in C(\ga_n)$, and
\eqref{eq:chain-length} holds. Therefore
$$
S_{DS} * \id (\Ga)
= S_{DS}(\Ga) - \sum_{\Cc \in C(\Ga)} (-1)^{l(\Cc)} \Om(\Cc)
= 0 = u \eps(\Ga);
$$
in other words, $S_{DS}$ is a left inverse for $\id$, and therefore
it is an antipode.
\end{proof}

As a corollary $S = S_B = S_{DS}$. Nevertheless, it is more instructive to
check that $S = S_{DS}$ \textit{identically}, as follows.

\begin{prop}
\label{pr:DS-antpbis}
$S_{DS}$ coincides with $S$ without cancellations.
\end{prop}

\begin{proof}
First, given a proper, connected graph $\Ga$, we rewrite
$$
S_{DS}(\Ga) := \sum_k (-1)^{k+1} \sum_{\Cc \in C_k(\Ga)} \Om(\Cc),
$$
where $C_k(\Ga)$ denote the set of chains of length $k+1$. Thus, it is
enough to prove that
$$
(-1)^{k+1} \sum_{\Cc \in C_k(\Ga)} \Om(\Cc) =
(u \eps -\id)^{*(k+1)}(\Ga) = \sg^{*(k+1)}(\Ga).
$$
To prove this first we notice that
\begin{equation}
\sum_{\emptyset \varsubsetneq \ga_1 \varsubsetneq \ga_2 \cdots \varsubsetneq
\ga_k}  \ga_1 \ox \ga_2/\ga_1 \oxyox \ga_k/\ga_{k-1} \ox \Ga/\ga_k
= \Dl'_k \cdots \Dl'_2 \Dl'(\Ga).
\label{eq:foldup}
\end{equation}
Indeed, by definition of the coproduct the statement is true for
$k = 1$. Moreover, if \eqref{eq:foldup} holds for $k-1$, then
\begin{align*}
\sum_{\emptyset \varsubsetneq \ga_1 \varsubsetneq \ga_2 \cdots \varsubsetneq
\ga_k \varsubsetneq \Ga}
&\ga_1 \ox \ga_2/\ga_1 \oxyox \ga_k/\ga_{k-1} \ox \Ga/\ga_k
\\
&= \sum_{\emptyset \varsubsetneq \ga_2 \varsubsetneq \ga_3
\cdots \varsubsetneq \ga_k  \varsubsetneq \Ga}
\Dl'_k(\ga_2 \ox \ga_3/\ga_2 \oxyox \ga_k/\ga_{k-1} \ox
\Ga/\ga_k)
\\
&= \Dl'_k \biggl( \sum_{\emptyset \varsubsetneq \ga_1 \varsubsetneq \ga_2
\cdots\varsubsetneq \ga_{k-1}  \varsubsetneq \Ga}
\ga_1 \ox \ga_2/\ga_1 \oxyox \ga_{k-1}/\ga_{k-2} \ox\Ga/\ga_{k-1} \biggr)
\\
&= \Dl'_k \Dl'_{k-1} \cdots \Dl'_2 \Dl'(\Ga).
\end{align*}
Thus, by \eqref{eq:many-conv-bis}
\begin{align*}
(-1)^{k+1} \sum_{\Cc \in C_k(\Ga)} \Om(\Cc)
&= (-1)^{k+1} \sum_{\emptyset \varsubsetneq \ga_1 \varsubsetneq \ga_2 \cdots
\varsubsetneq \ga_k  \varsubsetneq \Ga}
\ga_1 \, \bigl(\ga_2/\ga_1\bigr) \cdots
\bigl( \ga_k/\ga_{k-1} \bigr) \, \bigl(\Ga/\ga_k \bigr)
\\
&= (-1)^{k+1} m\,m_2 \cdots m_k\,\Dl'_k \cdots \Dl'_2 \Dl' (\Ga)
= \sg^{*(k+1)}(\Ga).
\qquad\qquad\qed
\end{align*}
\hideqed
\end{proof}

The proof shows how the chains are generated from the coproduct.

\smallskip

As an application, we obtain a nonrecursive formula for
$S_{T,f}$: if $\Cc = \set{\row{\ga}1k}$ is a chain in $C(\Ga)$, write
$$
\Om_{T,f}(\Cc) := T\biggl(T\biggl[ \cdots T\Bigl(T \Bigl[
T\bigl(f(\ga_1)\bigr) \, f(\ga_1/\ga_2)\Bigr] f(\ga_2/\ga_3)\Bigr)
\cdots f(\ga_{k-1}/\ga_k) \biggr] f(\Ga/\ga_k)\biggr).
$$
Let us use the temporary notation
$$
\tilde S_{T,f}\Ga :=
\sum_{\Cc \in C(\Ga)} (-1)^{l(\Cc)}\Om_{T,f}(\Cc)
= \sum_{k=0}^{\dl(\Ga)}
    (-1)^{k+1} \sum_{\Cc \in C_k(\Ga)} \Om_{T,f}(\Cc).
$$

\begin{prop}
\label{pr:twisted-DS}
$S_{T,f} = \tilde S_{T,f}$.
\end{prop}

\begin{proof}
We shall proceed by induction on the bidegree (no other method seems
available here). A simple check gives the statement for $\Ga \in \H^{(1)}$.
Assume the claim is true for graphs in $\H^{(l)}$ with $l \le k$, and let
$\Ga\in\H^{(k+1)}$, then
\begin{align*}
S_{T,f}\Ga
&= -[T\circ f]\Ga - T\Biggl[
\sum_{\emptyset\varsubsetneq\ga\varsubsetneq\Ga} S_{T,f}(\ga) f(\Ga/\ga)
\Biggr]
\\
&= -[T\circ f]\Ga - T\Biggl[
\sum_{\emptyset\varsubsetneq\ga\varsubsetneq\Ga} \sum_{\D \in C(\ga)}
(-1)^{l(\D)}\Om_{T,f}(\D)  f(\Ga/\ga) \Biggr],
\end{align*}
since each $\ga \in \H^{(l)}$ for some $l \le k$. Now, if $\D \in
C(\ga)$, then $\Cc = \D \cup \{\ga\} \in C(\Ga)$,
$$
l(\Cc) = l(\D) + 1, \sepword{and}\quad
\Om_{T,f}(\Cc) = T\bigl[\Om_{T,f}(\D) \, f(\Ga/\ga)\bigr].
$$
Conversely, if $\Cc = \{\row{\ga}1n\} \in C(\Ga)$ is not the trivial
chain $\{\emptyset\}$, then $\D = \{\row{\ga}1{n-1}\}$ is a chain in
$C(\ga_n)$, and
$$
(-1)^{l(\Cc)}\Om_{T,f}(\Cc)
= - T \bigl[ (-1)^{l(\D)}\Om_{T,f}(\D) f(\Ga/\ga_n)\bigr].
$$
Therefore
$$
S_{T,f} = \sum_{\Cc \in C(\Ga)} (-1)^{l(\Cc)}\Om_{T,f}(\Cc)
= \tilde S_{T,f}.
\eqno\qed
$$
\hideqed
\end{proof}

\smallskip

The morals of the story so far are: first, there is nothing in Bogoliubov's
procedure that will not be valid in \textit{any} connected, graded
bialgebra; second, Schmitt's formula for his incidence Hopf algebras
coincides identically with the geometric series
formulae~\eqref{eq:geom-series} and~\eqref{eq:many-conv}; third, the latter
in the field theory context gives rise to the Dyson--Salam formula. We turn
our attention now to Zimmermann's forest formula.

\section{Zimmermann's forest formula}

\begin{defn}
A (normal) \textit{forest} $\F$ of a proper, connected graph $\Ga$ is a
set of proper, divergent and connected subdiagrams, none of them equal to
$\Ga$, such that any pair of elements are nonoverlapping. Again we
include the forest $\set{\emptyset}$ as a special case. $F(\Ga)$ denotes
the set of forests of $\Ga$. The \textit{density} of a forest $\F$ is the
number $d(\F) = |\F| + 1$, where $|\F|$ is the number of elements of $\F$.
Given $\ga \in \F$ we say that $\ga'$ is a \textit{predecessor} of $\ga$ in
$\F$ if
$\ga'\varsubsetneq \ga$ and there is no element $\ga''$ in $\F$ such
that $\ga'\varsubsetneq \ga'' \varsubsetneq \ga$. Let
$$
\Th(\F) := \prod_{\ga \in \F\cup\{\Ga\}}  \ga / \tilde\ga,
$$
where $\tilde\ga$ denote the disjoint union of all predecessors of $\ga$.
When $\ga$ is minimal, $\tilde\ga = \emptyset$, and $\ga / \tilde\ga = \ga$.
\end{defn}

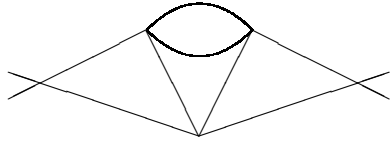
\begin{figure}[ht]
\begin{center}
\vspace{1pc}
\begin{picture}(0,50)
\put(0,0){\line(3,1){72}}
\put(0,0){\line(-3,1){72}}
\put(0,0){\line(1,2){20}}
\put(0,0){\line(-1,2){20}}
\put(20,40){\line(2,-1){52}}
\put(-20,40){\line(-2,-1){52}}
\qbezier(-20,40)(0,60)(20,40)
\qbezier(-20,40)(0,20)(20,40)
\end{picture}
\end{center}
\caption{Diagram $\Ga$ without extra cancellations in $S_Z(\Ga)$ with
respect to $S_{DS}(\Ga)$}
\label{fg:diagrama-prieto}
\vspace{1pc}
\end{figure}

Notice that if a forest $\F$ is a chain, then $\Th(\F) =\Om(\F)$, and
conversely if a chain $\Cc$ is a forest, $\Om(\Cc) = \Th(\Cc)$. Obviously
not every forest is a chain; but also not every chain is a forest, because
product subgraphs can occur in chains and cannot in forests. There are
diagrams like the one in the $\varphi^4_4$ model pictured in
Figure~\ref{fg:diagrama-prieto}, for which the sets of chains and forests
coincide; but in general there are fewer forests than chains.

Zimmermann's version for the antipode is defined by
$$
S_Z(\Ga) := \sum_{\F \in F(\Ga)} (-1)^{d(\F)} \Th(\F).
$$

\begin{prop}
\label{pr:Zimm-antp}
$S_Z$ provides another formula for the antipode of $\H$.
\end{prop}

\begin{proof}
Once more, the idea is to prove that $S_Z$ is an inverse, under
convolution, of $\id$. By definition
\begin{align*}
S_Z * \id (\Ga)
&= \sum_{\ga \subseteq \Ga} S_Z(\ga) \, \Ga/\ga
= S_Z(\Ga) + \sum_{\emptyset \subseteq \ga \varsubsetneq \Ga}
S_Z(\ga) \, \Ga/\ga
\\
&= S_Z(\Ga) + \sum_{\emptyset \subseteq \ga \varsubsetneq \Ga} \Bigl(
\prod_{i=1}^{n_\ga} S_Z(\la^\ga_i) \Bigr)\, \Ga/\ga
\\
&= S_Z(\Ga) + \sum_{\emptyset \subseteq \ga \varsubsetneq \Ga}
\prod_{i=1}^{n_\ga} \Bigl( \sum_{\F_i \in F(\la^\ga_i)}
(-1)^{d(\F_i)} \Th(\F_i) \Bigr) \, \Ga/\ga,
\end{align*}
where $n_\ga$ is the number of connected components of $\ga$, and
$\la^\ga_i$, $i= 1,\dots,n_\ga$ are the connected components of $\ga$.

Now, if $\F_i \in F(\la^\ga_i)$, then $\F := \bigcup_{i=1}^{n_\ga}
(\F_i \cup \set{\la^\ga_i})$ is a forest of $\Ga$. Moreover
$$
|\F| = \sum_{i=1}^{n_\ga} |\F_i| + n_\ga = \sum_{i=1}^{n_\ga} d(\F_i),
$$
so
\begin{equation}
d(\F) = \sum_{i=1}^{n_\ga} d(\F_i) + 1.
\label{eq:density}
\end{equation}
On the other hand, in $\F$, $\Ga/\tilde\Ga = \Ga/\ga$, hence
\begin{equation}
\Th(\F) =  \Bigl(\prod_{i=1}^{n_\ga} \Th(\F_i) \Bigr) \Ga/\tilde\Ga
= \Bigl(\prod_{i=1}^{n_\ga} \Th(\F_i) \Bigr) \, \Ga/\ga.
\label{eq:big-theta}
\end{equation}
Conversely, if $\F \in F(\Ga)$, and if $\row {\ga}1k$ are the maximal
elements of $\F$, then the sets $\F_i:= \set{\ga \in\F : \ga \varsubsetneq
\ga_i}$ constitute a forest of $\ga_i$. Since all the elements of a forest
are connected diagrams, $\row{\ga}1k$ are the connected components of
$\ga = \prod_{i=1}^k \ga_i$ and clearly \eqref{eq:density} and
\eqref{eq:big-theta} hold. Therefore,
$$
S_Z * \id (\Ga)
= S_Z(\Ga) - \sum_{\F \in F(\Ga)} (-1)^{d(\F)} \Th(\F)
= 0 = u \eps(\Ga).
$$
Thus, $S_Z$ is a left inverse for $\id$, and therefore is an antipode.
\end{proof}

The proofs of Propositions~\ref{pr:DS-antp} and~\ref{pr:Zimm-antp} are
parallel; the difference lies in minor combinatorial details. Even so, it is
clear that Zimmermann's formula (although more sensitive in practice to the
details of the renormalization method) is, from the combinatorial viewpoint,
altogether subtler than Bogoliubov's or Dyson and Salam's. It is more
economical in that all the cancellations implicit in the convolution
formula~\eqref{eq:geom-series} are taken into account and suppressed; this we
already made clear in the context of the algebra of rooted
trees.${}^{1}$ Thus, the ``commerce'' between quantum field
theory and Hopf algebra theory has not been one-way: Zimmermann's formula is
advantageoulsy applicable to a large class of bialgebras.

The reader will have no difficulty in writing the nonrecursive forest
formula for the twisted coinverse $S_{T,f}$.

\section{The bidegree and computations in quantum field theory}

Let us indicate first that Kreimer has announced${}^{17}$ a new
proof of finiteness of the renormalized graphs and Green functions, based
on a cohomological reinterpretation of the basic coproduct
equation~\eqref{eq:coprod}.

Whether the Connes--Kreimer algebraic paradigm will become useful to simplify
\textit{computations} ---in distinction to ``merely'' proofs--- in realistic
field theories, seems to hinge to a large extent on the practical usefulness
of the depth bigrading. The fact is that simplifications in sums of Feynman
diagrams do occur, and they usually involve, beyond trivialization of the
topology, reduction in depth. For instance, in the $\varphi^4_4$ model we have
(for the corresponding amplitudes in configuration space) the situation
described in Figure~\ref{fg:magic-sum}.

\begin{figure}[h]
\begin{center}
\vspace{2pc}
\parbox{2pc}{
\begin{picture}(0,0)
\put(-10,-15){\line(0,1){30}}
\put(-10,-15){\line(2,3){20}}
\put(10,-15){\line(0,1){30}}
\put(10,-15){\line(1,0){8}}
\put(-10,15){\line(-1,0){8}}
\qbezier(-10,-15)(0,-25)(10,-15)
\qbezier(-10,-15)(0,-5)(10,-15)
\qbezier(-10,15)(0,5)(10,15)
\qbezier(-10,15)(0,25)(10,15)
\end{picture}
}
+
\qquad\quad
\parbox{2pc}{
\begin{picture}(0,0)
\put(0,-20){\line(1,1){20}}
\put(0,-20){\line(-1,1){20}}
\put(-26,0){\line(1,0){52}}
\qbezier(0,0)(10,-10)(0,-20)
\qbezier(0,0)(-10,-10)(0,-20)
\qbezier(-20,0)(0,20)(20,0)
\end{picture}
}\qquad
=
\hspace{4.5em}
\parbox{4pc}{
\begin{picture}(0,0)
\put(30,0){\line(1,0){6}}
\put(-30,0){\line(-1,0){6}}
\qbezier(-30,0)(-20,10)(-10,0)
\qbezier(-30,0)(-20,-10)(-10,0)
\qbezier(-10,0)(0,10)(10,0)
\qbezier(-10,0)(0,-10)(10,0)
\qbezier(10,0)(20,10)(30,0)
\qbezier(10,0)(20,-10)(30,0)
\qbezier(-30,0)(0,40)(30,0)
\end{picture}
}
+ \ $\displaystyle\frac{1}{2}$
\hspace{2.5em}
\parbox{3pc}{
\begin{picture}(0,0)
\put(20,0){\line(1,0){6}}
\put(-20,0){\line(-1,0){6}}
\qbezier(-20,0)(-10,10)(0,0)
\qbezier(-20,0)(-10,-10)(0,0)
\qbezier(0,0)(10,10)(20,0)
\qbezier(0,0)(10,-10)(20,0)
\qbezier(-20,0)(0,30)(20,0)
\end{picture}
}
\end{center}
\vspace{1pc}
\caption{$\zeta(3)$ coefficients vanish in the sum of two graphs with
the same symmetry factor}
\label{fg:magic-sum}
\vspace{1pc}
\end{figure}

A naive hope in that respect, to wit, that every diagram be eventually
expressed in terms of primitive elements (so renormalization proceeds ``at a
stroke'') is quickly dashed. A Hopf algebra $H$ is
\textit{primitively generated} when the smallest subalgebra of $H$ containing
all its primitive elements is $H$ itself. The structure theorem for
commutative connected graded algebras${}^{21}$ makes it plain that Hopf
algebras of Feynman graphs are far from being primitively generated; neither
the Hopf algebra of rooted trees nor its noncommutative geometry subalgebra
$H_{\mathrm{CM}}$ (Ref.~22) are primitively generated.

In fact, only elements for which the coproduct is invariant under the flip map
$a \ox b \mapsto b \ox a$ can be primitively generated. In any Hopf algebra
associated to a field theory there exist ``ladder'' subalgebras of diagrams
with only completely nested subgraphs, and these subalgebras are primitively
generated. However, this is of scant practical use, as then the recursive
methods${}^{23}$ carry off the award for computational
simplicity.

According to the structure theorem, commutative Hopf algebras can be
decomposed as algebras as a tensor product
$$
H = S(P(H)) \ox S(W_H),
$$
where $S(P(H))$ denotes the polynomial algebra generated by all the primitive
elements in $H$, and $S(W_H)$ the polynomial algebra on a (nonunique) suitable
subspace $W_H$ of $H$. To get a handle on (a representative for)
$W_H$ for bialgebras of Feynman graphs is on the order of the day.

The depth grading for the algebra $H_R$ of rooted trees has been investigated,
beyond Ref.~13, in Ref.~24. The strategy suggested in Ref.~24 looks feasible
in bialgebras of graphs. By use of the dual algebra, so-called normal
coordinate elements (appropriate sums of products of graphs) can be found, for
which the antipode (although not the twisted antipode in general) is
\textit{diagonal}.

Let $Z_\ga$ be the dual element of a graph $\ga$. Any graph $\Ga$ has an
associated normal element $\psi_\Ga$, and any graph can be decomposed
into a sum of products of normal elements. Let a sequence of graphs $J =
(\ga_1,\dots,\ga_k)$ be given; we say that $\Ga$ is compatible with $J$ if
\begin{equation}
\<Z_{\ga_1}\ox\dots\ox Z_{\ga_k},\Dl^{k-1}\Ga> \neq 0.
\label{eq:compatibility}
\end{equation}
The normal decomposition is as follows:
$$
\Ga = \psi_\Ga +
\sum_{k\geq2}\frac{\<Z_{\ga_1}\ox\dots\ox Z_{\ga_k},\Dl^{k-1}\Ga>}{k!}\,
\psi_{\ga_1}\dots\psi_{\ga_k},
$$
where we sum in practice over a finite number of compatible sequences.

Ladder normal coordinate elements are primitive, and for non-ladder ones
indications are that the complexity of their renormalization is substantially
lessened with respect to that of the graphs themselves; they are instrumental
in the description of~$W_H$.

Also, Kreimer has introduced${}^{17}$ a ``shuffle" product of diagrams, based
on a variant of~\eqref{eq:compatibility}, that seems to hold promise of
eventual factorization of perturbative field theory into primitive elements.
In this respect, as in other tantalizing subjects springing from the
Connes--Kreimer paradigm, we are barely starting to scratch the surface.

\section*{Acknowledgements}

We thank Joseph C. V\'arilly for illuminating discussions and
\TeX{}nical help. We are grateful for the hospitality of the Departamento
de F\'{\i}sica Te\'orica of the Universidad de Zaragoza. Support from the
Vicerrector\'{\i}a de Investigaci\'on of the Universidad de Costa Rica is
acknowledged.

\appendix

\section{Appendix: Graded bialgebras}

A bialgebra H is a vector space over a field $\FF$ (here taken to be of
characteristic~0) equipped with two structures: an algebra structure
and a coalgebra structure, related by some compatibility conditions.
The algebra structure is described by two maps: the product
$m \: H \ox H \to H$, and the unit map $u \: \FF \to H$. The
conditions imposed on these maps are:
\begin{enumerate}
\item
Associativity: $m(m \ox \id) = m(\id \ox m) : H \ox H \ox H \to H$;
\item
Unity: $m(u \ox \id) = m(\id \ox u) = \id : H \to H$.
\end{enumerate}

A coalgebra is obtained by reversing arrows in the defining maps for an
algebra; it is, therefore, also described by two maps: the coproduct
$\Dl \: H \to H \ox H$, and the counit $\eps \: H \to \FF$. The
requirements are:
\begin{enumerate}
\addtocounter{enumi}{2}
\item
Coassociativity: $(\Dl \ox \id)\Dl = (\id \ox \Dl)\Dl : H \to H \ox H
\ox H$;
\item
Counity: $(\eps \ox \id) \Dl = (\id \ox \eps) \Dl = \id : H \to H$.
\end{enumerate}

Finally, to obtain a bialgebra one stipulates
\begin{enumerate}
\addtocounter{enumi}{4}
\item
Compatibility: $\Dl$ and $\eps$ are unital algebra homomorphisms.
\end{enumerate}

This requirement turns out to be equivalent to asking that $m$ and
$u$ be coalgebra morphisms.

\begin{defn}
A bialgebra $H = \bigoplus^\infty_{n=0} H^{(n)}$ graded as a vector
space is called a \textit{graded bialgebra} when the grading is
compatible with both the algebra and the coalgebra structures:
$$
H^{(n)}H^{(m)} \subseteq H^{(n+m)} \sepword{and}\quad
\Dl(H^{(n)}) \subseteq \bigoplus_{p+q = n} H^{(p)} \ox H^{(q)}.
$$
It is called \textit{connected} when the first piece consists of
scalars only: $H^{(0)} = \FF\, 1$.
\end{defn}

A most useful property of connected graded bialgebras is that when
$a \in H^{(n)}$, the coproduct can be written as
\begin{subequations}
\label{eq:gr-coprod}
\begin{equation}
\Dl a = a \ox 1 + 1 \ox a + \sum_j a'_j \ox  a''_j,
\label{eq:gr-coprod-full}
\end{equation}
where the elements $a'_j$ and $a''_j$ all have degree between 1 and $n-1$. The
proof is easy and found in many places.${}^{11,15,25,26,27}$ To simplify the
notation we define
\begin{equation}
\Dl'a := \Dl a - a \ox 1 - 1 \ox a = \sum_j a'_j \ox  a''_j.
\label{eq:gr-coprod-trunc}
\end{equation}
\end{subequations}
Coassociativity of $\Dl'$ is easily obtained from the coassociativity
of $\Dl$. An element $a \in H$ is called (1-)primitive when $\Dl'a = 0$.

Applying $\eps \ox \id$ to \eqref{eq:gr-coprod-full} gives
$a = (\eps \ox \id)(\Dl a) = \eps(a)1 + a + \sum_j \eps(a'_j)\,a''_j$;
therefore, if $a \in H^{(n)}$ with $n \geq 1$, the connectedness
condition forces $\eps(a) = 0$.

\bigskip\bigskip

{\bf References}

\bigskip

1. 
H. Figueroa and J. M. Gracia-Bond\'{\i}a,
\textit{Mod. Phys. Lett.} {\bf A16}, 1427 (2001).

2.
A. Connes and D. Kreimer,
\textit{Commun. Math. Phys.} {\bf 199}, 203 (1998).

3.
W. Zimmermann,
\textit{Commun. Math. Phys.} {\bf 15}, 208 (1969).

4.
E. B. Manoukian,
\textit{Renormalization},
(Academic Press, London, 1983).

5.
A. Connes and D. Kreimer,
\textit{Commun. Math. Phys.} {\bf 210}, 249 (2000).

6.
J. M. Gracia-Bond\'{\i}a and S. Lazzarini,
Connes--Kreimer--Epstein--Glaser renormalization,
hep-th/0006106.

7.
J. M. Gracia-Bond\'{\i}a,
\textit{Math. Phys. Analysis and Geometry} {\bf 6}, 59 (2003).

8.
W. Zimmermann,
Remarks on equivalent formulations for Bogoliubov's method of
renormalization, in \textit{Renormalization Theory}, eds. G. Velo and A. S.
Wightman, NATO ASI Series C~23 (D. Reidel, Dordrecht, 1976).

9.
D. Kastler,
On the external structure of graphs, to appear in \textit{J. Math.
Phys.}

10.
G. Pinter,
\textit{Lett. Math. Phys.} {\bf 54}, 227 (2000).

11.
J. M. Gracia-Bond\'{\i}a, J. C. V\'arilly and H. Figueroa,
\textit{Elements of Noncommutative Geometry} (Birkh\"auser, Boston, 2001).

12.
D. Kreimer,
\textit{Adv. Theor. Math. Phys.} {\bf 2}, 303 (1998).

13.
D. J. Broadhurst and D. Kreimer,
\textit{Commun. Math. Phys.} {\bf 215}, 217 (2000).

14.
H. Kleinert and V. Schulte-Frohlinde,
\textit{Critical Properties of $\phi^4$ Theories}
(World Scientific, Singapore, 2001).

15.
J. C. V\'arilly,
Hopf algebras in noncommutative geometry, in
\textit{Geometrical and Topological Methods in Quantum Field Theory},
eds. A. Cardona, H. Ocampo and S. Paycha
(World Scientific, Singapore, 2003).

16.
D. Kreimer,
\textit{Adv. Theor. Math. Phys.} {\bf 3}, 3 (1999).

17.
D. Kreimer,
\textit{Ann. Phys.} {\bf303}, 179 (2003).

18.
K. Ebrahimi-Fard, L. Guo and D. Kreimer,
Integrable renormalization II: the general case,
hep-th/0403118.

19.
A. Connes and D. Kreimer,
\textit{Commun. Math. Phys.} {\bf 216}, 215 (2001).

20.
W. R. Schmitt,
\textit{J. Comb. Theory} {\bf A46}, 264 (1987);
\textit{J. Pure Appl. Alg.} {\bf 96}, 299 (1994).

21.
K.-H. Hoffmann and S. A. Morris,
\textit{The Structure of Compact Groups}
(de~Gruyter, Berlin, 1998).

22.
A. Connes and H. Moscovici,
\textit{Commun. Math. Phys.} {\bf 198}, 198 (1998).

23.
I. Bierenbaum,
\textit{Die Riemannsche $\zeta$-Funktion in iterierten
Einschleifenintegralen} (Diplomarbeit, Mainz, 2000).

24.
C. Chryssomalakos, H. Quevedo, M. Rosenbaum and J. D. Vergara,
\textit{Commun. Math. Phys.} {\bf 225}, 465 (2002).

25.
N. Bourbaki,
\textit{Groupes et Alg\`ebres de Lie}
(Hermann, Paris, 1972).

26.
D. Kastler,
Connes--Moscovici--Kreimer Hopf algebras,
in \textit{Mathematical Physics in Mathematics and Physics:
Quantum and Operator Algebraic Aspects}, ed. R. Longo,
Fields Institute Communications~30 (American Mathematical Society, Providence,
RI, 2001).

27.
S. Montgomery,
\textit{Hopf Algebras and their Actions on Rings},
CBMS Regional Conference Series in Mathematics~82,
(American Mathematical Society, Providence, RI, 1993).

\end{document}